# Investigation of the Dynamics of Liquid Cooling of 3D ICs

Sakib Islam
*Department of Electrical Engineering*
Northern Illinois University
Dekalb, USA
aislam@niu.edu

Ibrahim Abdel-Motaleb[*]
*Department 0f Electrical Engineering*
Northern Illinois University
Dekalb, USA
ibrahim@niu.edu

*Abstract*—Although 3D IC technology can provide very high integration density, they suffer from having hotspots that may reach 1000's of degrees. To manage this heat, it is necessary to study the dynamics of cooling and thermal behavior of the ICs. In this study, we report on the dynamics of microchannel liquid cooling using water, R22, and liquid nitrogen. The study shows that using diamond cooling blocks ensures normal operating temperature of 60 °C or less, using any of the above coolants. Using $SiO_2$ blocks, only liquid nitrogen can provide acceptable operating temperatures. The study shows also that liquid latent energy and inlet velocity play a major role in the cooling dynamics.

## I. Introduction

3D ICs provide a way to achieve the high integration density craved for by industry. But such high density produces local hot spots with temperatures reaching thousands of kelvins. Hot spots may cause temperature variations across the chip that may affect the overall performance of the IC. Hot spots may also affect the resistance of the interconnects slowing the communication between the different parts of the IC. This situation gets even more critical, since according to the International Technology Roadmap for Semiconductors (ITRS), for a single package, the power density may need to increase to up to $10^6$ W/cm$^2$ to meet future applications [1]. To address these issues, thermal management techniques need to be developed to ensure that no hot spot can exceed the IC operating temperature.

Conventional 2D IC cooling methods employ heat sinks on the face of the IC to transfer heat from the ICs to the outside environment. Because the power dissipation is not excessively high, this technique can maintain chip temperature to within the operating temperature range. In the case of the vertically stacked 3D IC chips, there are multiple heated surfaces, and it is almost impossible to place those bulky heat sinks on top of each of the stacked layers. Cooling from the surface would not be as efficient, since heat coming from the inside layers will have a long heat path that complicates the cooling process, unless fast heat removal techniques can be employed [2].

To solve the cooling issues of the 3D ICs, other cooling techniques have been proposed. Among them are Microelectromechanical Systems (MEMS)-based technology, embedded micro channels [3], liquid immersion cooling [4], and microfluidic cooling using thermal TSVs [5].

In this study, we report the results of investigating the dynamics of cooling a 1000 W/cm$^2$ local hot spot, using $SiO_2$ and diamond cooling blocks and employing water, Liquid Nitrogen (LN), and R22 as coolants. The study was conducted using advanced multi-physics numerical analysis program, Comsol$^{TM}$,

## II. Heat transfer mechanism

Heat transfer between the cooling blocks and the 3D IC can be due to a combination of conduction, convection and radiation. Heat conduction describes the flow of heat through materials. In general, conduction is governed by,

$$\frac{Q}{t} = KA\frac{\Delta T}{l} \quad (1)$$

Here, $Q$ is the heat energy, $t$ is the time, $k$ is the thermal conductivity (W/mK), $A$ is the cross-sectional area of the flowing path, $l$ is the length of the flowing path, and $\Delta T$ is the temperature difference between the beginning and end of the path in Kelvins (K).

Convection heat flow is a result of the transport of heat by the motion of gas or fluid molecules. Heat convection can be described by the following equation,

$$Q = hA\Delta T \quad (2)$$

In this equation, $h$ is the convective heat transfer coefficient (J/m$^2$K) [6], $A$ is the object's cross sectional area that carries the fluid, and $\Delta T$ is the temperature difference between fluid and heated surface.

Radiation is the emission of heat from hot body to the ambient. Radiation power can be obtained from the relation,

$$P = \varepsilon\sigma AT^4 \quad (3)$$

Here, $\varepsilon$ is the emissivity factor, $\sigma$ is Stefan-Boltzmann constant (=1.38X10$^{-23}$ m$^2$kg/s$^2$ K), $A$ is the surface area of the radiating body and $T$ is the temperature.

## III. Design of the cooling system

The 3D IC structure is represented by a 1000x2000 μm$^2$ tungsten heater on 5x5mm$^2$ Si wafer. The heater is made of 200 μm Si substrate with a 3000 Å of $SiO_2$ on its top. Next, a 0.18 μm serpentine-shaped tungsten-heater is deposited and covered with 1500 Å of $SiO_2$. Aluminum pads of an area of 125x250 μm$^2$ and thickness of 0.05 μm are deposited and connected to the ends of the heater using vias in the $SiO_2$ layer. Finally, a 7000 Å $SiO_2$ layer is deposited for passivation. The area of the silicon chip is 5mmx5mm and the W-heater has the shape shown in Fig. 1.

Fig. 1 shows the heater structure on top of the silicon wafer. The widths of the wire of the heater is 120 μm and the separation between the wires is 115 μm. The aluminum pads are shown at the ends.

Fig 2. Shows the cooling block, where it has two microchannels with the heater chip sandwiched between them. Both microchannels length and width are 40 mm and the opening is 2x40 mm². The walls of the microchannels are 250 μm thick. The liquid enters from the inlet and exits from the other side.

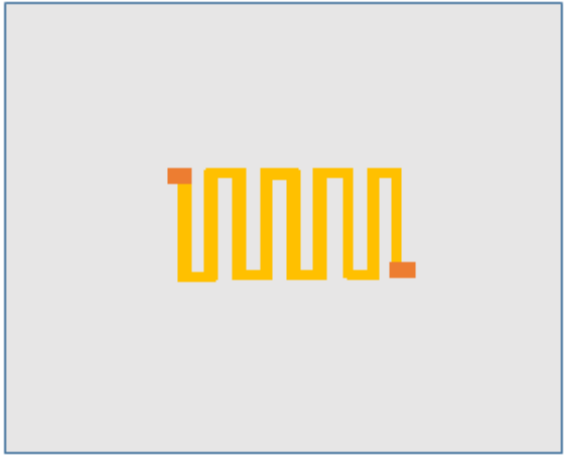

Fig.1: The serpentine shaped tungsten heating block with aluminum contacts at the two ends.

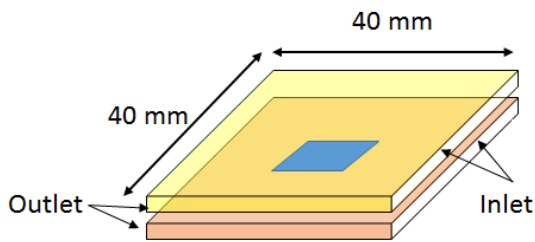

Fig. 2. The geometry of the cooling block, with the chip inbetween the lower and upper channels. Channel dimensions are 40mm x 40 mm x 2mm; wall thickness is 250 μm.

Two cooling blocks are used: the first is made of $SiO_2$ and the second of diamond. Three liquid coolants were used: water, R22, and Liquid Nitrogen (LN). The size of the block is made large to ensure laminar flow of the coolant at the inlet. Turbulent flow is left for another study.

## IV. SIMULATION AND ANALYSIS

Fig. 3, shows Comsol meshing structure for the cooling blocks, where the size of the mesh decreases at the edge to ensure accurate simulation. In this simulation, we used emissivity value of 0.79 for $SiO_2$ [6] and 0.63 for diamond [7]; the thermal conductivity used are 1.4 W/m.K for $SiO_2$, 990 W/m.K for diamond, 130 W/m.K for Si, 174 W/m.K for tungsten and 238 W/m.K for aluminum. The convective heat transfer coefficient is assumed to be 50 W/m²K for all the coolants and 10 W/m²K for air.

The hot spot temperature was simulated without coolant, with just natural air convection and radiation. The results show that with 20W, temperature reached 4350 K or 4077 °C. In reality, the chip would be evaporated before reaching this temperature.

Next, temperature was simulated when liquid coolants were admitted. The hot spot temperature was then monitored for 15s. The heater power is controlled by the input voltage. Here, the maximum power used is 20 W, and this resulted in 1000 W/cm² hot spot, or 80W/cm² across the 5mmx5mm chip. The simulation was performed for the $SiO_2$ and diamond cooling blocks. For each block, water, R22, and LN are used.

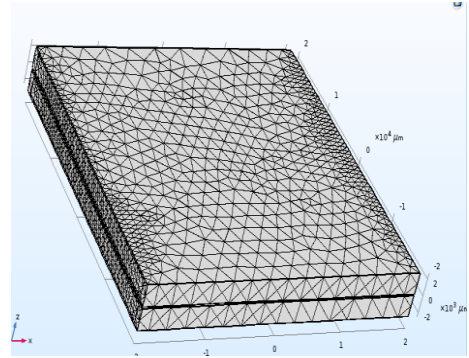

Fig.3: Comsol mesh of the geometry. Size of the meshes change to ensure accurate simulation results.

### A. SILICON DIOXIDE COOLING BLOCK

The thermal analysis of the $SiO_2$ cooling block was studied, where colants were admitted at the same time the heater was powered. The initial temperature of all regions and materials, including the ambient, are set to the room temperature, 293 K. Inlet velocities of 10 mm/s and 100 mm/s were used to investigate the effect of the velocity on the dynamics of the cooling; see Fig. 4 and Fig. 5, respectively.

For 10 mm/s velocity, it was noticed that the hot spot temperature overshoots. for LN, temperature overshooted to 375 K before setlling to 293 K. For R22, temperature increased to 438K, before it decreased to a constant value of 414K. For water, the temperature increased to to 436 before saturating to 434.

This results appears to be for a single phase cooling, where the cooling is done using liquid. If we have two phase cooling, where both liquid and gas contribute to the cooling proces, we expect to have lower temperature. This in fact what was noticed in [8]. This decrease in temperature can be attributed to the high latent energy of liquid, or to the high energy required to evaporate the liquid. It is known that water has high latent energy of 2230 J/gm. For R22, the latent energy is233 J/gm and for LN it is 200 J/gm. To convert liquid to gas, water needs 10 time the energy needed for R22 or LN. Therefore, water cooling may be more efficient but to reach the stage of evaportion, temperature should increase to 100 ˚C, which may not be acceptable for IC operation.

When the velocity increased to 100 mm/s, overshoots are almost eliminated; temperature rises only from 293 K to 304 K. For LN, the temperature is reduced to a saturation value of 207K in one second. For R22, the temperatue did not decrease; instead, it increased to 362 K. For water, the temperature staurated at 416 K. As can be seen, cooling using water or R22 is not acceptable, since the saturation temperatures are higher than the maximum acceptable operating temperature of 60˚C, or 333K. This, however, can be solved by doing the following: (a) admitting the liquid colant before powering the circuit and (b) using high thermal conductivity materials, such as diamond, to build the the cooling block.

The diamond cooling block was next simulated for inlet velocity of 10 mm/s and 100 mm/s. The results show that the maximum (overshoot or saturation) temperature did not exceed 333K, or 60 ˚C, for each of the three coolants at any of the velocities. The enhanced cooling effeciency is attributed to the high thermal conductivety of diamond than $SiO_2$. This means all three coolants can be used to provide the required operating temperature.

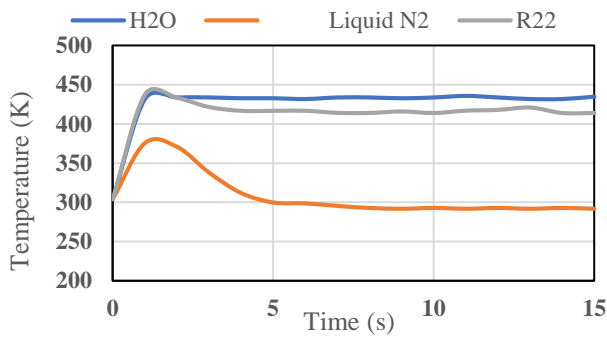

Fig. 4. Hot spot temperature for water, LN, and R22 coolant with 10 mm/s inlet velocity.

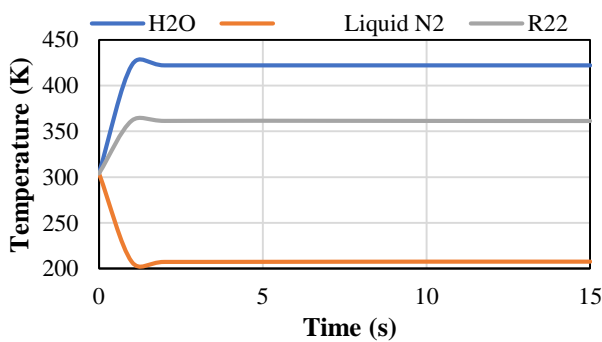

Fig. 5. Hot spot temperature for water, LN, and R22 coolant with 100 mm/s inlet velocity.

It appears that the reason for the overshoot, shown in Fig. 4, is due to the powering of the chip at the same time the coolant is admitted. In this case, temperature increaseed fast because the initial temperature of the block and the ambient were assumed to be 293K. To detrmine if this assumption is correct or not, collant was admitted first until the temperature of the block and the chip reached the coolant temperature which is 293K for water, 232.2 K for R22, and 77K for LN. The ambient temperature outside the block remained at room temperature.

The results shown in Fig. 6 confirm that the overshoot has been eliminated for inlet velocity of 10 mm/s. The behavior remains the same when the velocity is increased to 100 mm/s. In this case, for 10 mm/s velocity, a staruration temperature of 123 K, 272K, and 325 K, for LN, R22, and $H_2O$, respectively, were reached. But, as shown in Fig. 7, these temperatures increased to 106 K, 260 K, and 318 K from 77, 232.2, and 293 K, respectively, when the velocity increased to 100 mm/s. Therefore, it is believed that high conductivity materials, such as diamond, can provide cooling systems that ensure acceptable operating temperatures for 3D ICs.

## V. CONCLUSION

We designed cooling blocks to study the dynamics of liquid cooling of 3D ICs. The results shows that an overshoot takes place, if chip powering and colant admission are done at the same time. The results also show that using high thermal conductivity material enhances the cooling process.

To further undersatnd the dynamics of the cooling process, the internal pressure and velocity of the microchannels should be investigated. To evaluate the potential of product failure due to the cooling cycles, the stress and strain of the chip may be studied. To reduce cost, one would experiment with cooling blocks made of SiC, aluminum, or steel.

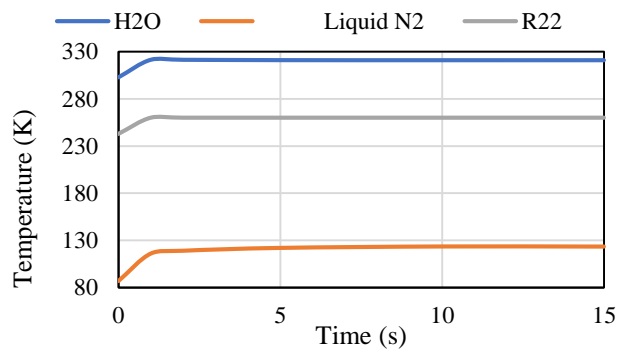

Fig. 6. Hot spot temeprature for diamond block when coolant is admitted before powering the chip. Inlet velocity is 10 mm/s.

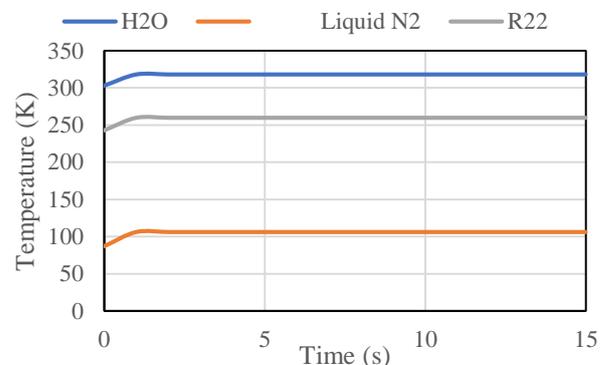

Fig. 7. Hot spot temeprature for diamond block when coolant is admitted before powering the chip. Inlet velocity is 100 mm/s.